\documentclass{cjaa}
\usepackage{graphicx,times,natbib,amssymb}
\bibpunct[, ]{(}{)}{;}{a}{}{,}
\begin{document}
\title{A Critical Review of the Evidence for M32 being a Compact Dwarf Satellite of M31 rather than a More Distant Normal Galaxy}

\volnopage{Vol.\ 8 (2008), No. 4,~ 369--384}
   \setcounter{page}{369}

\author{C.\ Ke-shih Young
         \inst{1,2,3,4}
        \and Malcolm J.\ Currie
         \inst{5}
        \and Robert J.\ Dickens
         \inst{5}
        \and A-Li Luo
         \inst{1}
        \and Tong-Jie Zhang
         \inst{6}
}

\institute{
National Astronomical Observatories, Chinese Academy of Sciences, Beijing 100012, China\\
\and
Shanghai Astronomical Observatory, Chinese Academy of Sciences, Shanghai 200030, China\\
\and
Department of Physics, University of Hong Kong, Hong Kong, China\\
\and
School of Physics, University of New South Wales, Sydney 2052, Australia; {\it cky@bat.phys.unsw.edu.au}\\
\and
Rutherford Appleton Laboratory, Chilton, Didcot, Oxfordshire OX11 0QX, Great Britain\\
\and
Department of Astronomy, Beijing Normal University, Beijing 100875, China \\
\vs\no
   {\small Received 2007 October 1; accepted 2008 January 16}
}

\abstract{ Since Baade's photographic study of M32 in the  mid
1940s, it has been accepted as an established fact that M32 is a
compact dwarf satellite of M31. The purpose of this paper is to
report on the findings of our investigation into the nature of the
existing evidence. We find that the case for M32 being a satellite
of M31 rests upon Hubble Space Telescope (HST) based stellar
population studies which have resolved  red-giant branch (RGB) and
red clump stars in M32 as well as other nearby galaxies. Taken in
isolation, this recent evidence could be considered to be
conclusive in favour of the existing view. However, the
conventional scenario does not explain M32's anomalously high
central velocity dispersion for a dwarf galaxy (several times that
of either NGC 147, NGC 185 or NGC 205) or  existing planetary
nebula observations (which suggest that M32 is more than twice as
distant as M31) and also requires an elaborate physical
explanation for M32's inferred compactness. Conversely, we find
that the case for M32 being a normal galaxy, of the order of three
times as distant as M31, is supported by: (1) a central velocity
dispersion typical of intermediate galaxies, (2) the published
planetary nebula observations, and (3) known scaling relationships
for normal early-type galaxies. However, this novel scenario
cannot account for the high apparent luminosities of the RGB stars
resolved in the M32 direction by HST observations. We are
therefore left with two apparently irreconcilable scenarios, only
one of which can be correct, but both of which suffer from
potentially fatal evidence to the contrary. This suggests that
current understanding of some relevant fields is still very far
from adequate.
\keywords{ galaxies:
individual: M32 --- galaxies: distances and redshifts
--- galaxies: dwarf --- galaxies: elliptical and lenticular, cD
--- galaxies: fundamental parameters
} }

\authorrunning{C.K.S.\ Young et al.}
\titlerunning{On the Evidence for M32 being a Compact Dwarf Galaxy}
\maketitle

\section{Introduction}
\label{Sec1}

As M32 appears projected onto the disc of M31, separation of the
two galaxies' integrated light and constituent stars is
notoriously difficult. Consequently, hardly any formal distance
measurements are available for M32. On the other hand, much effort
has gone into attempting to demonstrate the existence of physical
associations between M31 and M32 (which would imply physical
proximity) and/or that M32 is in front of M31's disc (which would
place a hard upper limit on M32's distance). This is because since
\citet{B 1944} it has been generally accepted that M32 must be a
satellite of M31. M32's distance has therefore not been regarded
as a major issue as it has generally been assumed to be similar to
M31's (and therefore to be known at least approximately). Note
though that before \citet{B 1944} there was no consensus on M32's
status with respect to M31 or on its distance.

Assuming that M32 is at a similar distance to that of M31 (whose
distance we take to be 0.76~Mpc from e.g.\ \citealt{Vdb 2000a})
then, on account of its angular size and degree of central
concentration, it must be an unusually compact dwarf
galaxy--regardless of  whether it is elliptical or lenticular in
nature. Note that although M32 has long been presumed to be an
elliptical galaxy, the possibility that it
may instead be a lenticular needs to be taken seriously
\citep{G 2002}. Since e.g.\ \citet{Dv 1961}, it has been accepted
that M32 is a red compact galaxy (RCG). Most members of this
extremely rare class of galaxy appear to be associated with larger
neighbours and since \citet{K 1962} it has been accepted that
their compactness arises due to tidal truncation by their
neighbours. Red compact dwarfs (RCDs) are all the more unusual
(and rare) because their red intrinsic colours and high degree of
central concentration with respect to mass are characteristic of
early-type giants and intermediates, not dwarfs. In addition to
M32, only five other candidates have been identified
\citep{C+ 2007}. However, as is evident from figures~3 and 4 of
\citet{C+ 2007}, M32 remains the faintest, smallest and most
compact RCD [candidate] known, and therefore the most extreme case
in terms of its deviation from most known scaling relationships
for normal galaxies.

In Appendix~\ref{AppA}, in a radical departure from current
practices, we treat M32 as if it were a normal galaxy in order to
investigate what distance it would need to be projected to in
order for it to fit as many scaling relationships as possible. We
find that if M32 were projected to a distance of 2.3($\pm$0.8)
[$-$0.77$A_B$(M31)~mag$^{-1}$]~Mpc and if $A_B$(M31) ($B$-band
absorption due to M31) were small, it would obey the known scaling
relationships defined by normal early-types, and a physical
explanation for its inferred extreme compactness would not be
necessary. With this result in mind, in this paper, we test our
radical new hypothesis against the existing evidence, the null
hypothesis being that M32 is an RCD satellite of M31.

Unfortunately, in the testing of any controversial new hypothesis
a critical approach is often unavoidable, and this paper is no
exception. It is not our aim to be unnecessarily critical of the
existing results, but as we hope most readers will agree, we
believe that it is a necessary exercise to investigate the
possibilities for re-interpreting evidence that runs contrary to
the new hypothesis. Only this way can robust evidence (that is not
open to re-interpretation) be separated from circumstantial
arguments (that are). Having said this, our aim is to find the
truth and we do not have a vested interest in the outcome. This
paper merely documents a thought experiment that we have
conducted. However, it may also be of historical interest and even
shed some light on how science is done.

In Sections~\ref{Sec2}, \ref{Sec3} and \ref{Sec4}, italic
type is used for the existing evidence in order to distinguish
it from our own analysis. It was found to be impractical to
present the existing evidence in strict chronological order.
Instead, we have attempted to introduce readers to the issues
in an order that requires minimal prior knowledge of the subject,
whilst at the same time minimizing the need for any repetition.
With a view to remaining objective throughout this paper, we have
adopted wordings that neither imply that M32 is an RCD nor imply
that it is a background galaxy.

\section{Distance-related issues}
\label{Sec2}

\subsection{Evidence of association with M31}
\label{Subsec2.1}

$\bullet$ {\sl The centroid-to-centroid distance on the sky for
M31 and M32 is a mere 24~arcmin (or a separation tangential to the
line of sight of 5.3~kpc at a distance of 0.76~Mpc). Also, the
heliocentric radial velocities of the two galaxies are reasonably
similar: $-$205~km~s$^{-1}$ for M32 cf.\ $-$301~km~s$^{-1}$ for
M31. This suggests that they are neighbours.}

This piece of evidence is of course circumstantial and does not
exclude the possibility that the Milky Way, M31 and M32 might
actually be in approximate geometrical alignment, with M32 lying
up to a few Mpc to the rear of M31. Furthermore, if such an
approximate alignment existed, it would not necessarily need to be
a chance alignment. Instead it might be a product of the structure
of the Local Group itself.

$\bullet$ {\sl Since \citet{K 1962}, it has been the common wisdom
that M32 is tidally truncated by M31, thereby requiring that the
two galaxies be very close neighbours.}

The deepest luminosity profiles of M32 available are those of
\citet{K 1987} and \citet*{CGJ 2002}. In spite of the differences
in photometric pass band and isophote-fitting procedures, as noted
by \citet{G 2002} these profiles are qualitatively very similar to
one another. Both profile datasets exhibit a luminosity excess (with
respect to a single-component \citealt{S 1968} model) at radial
distances ($r$) of $\sim$200 to 250~arcsec and both exhibit a very
minor downturn (with respect to both one and two-component models)
at $r \sim 250$~arcsec \citep{G 2002}.

That \citet{K 1987} found ``no evidence that the profile is truncated'',
whilst \citet{CGJ 2002} followed their mention of ``a subtle downturn ...
at $r \sim 250$~arcsec'' with a note of caution, is a clear indication
as to how ambiguous the existing observations remain. In galaxy surface
photometry, reliable `background' subtraction is often difficult at the
best of times, but in the case of M32, the difficulties are extreme due
to the need to remove the irregular `background' light contribution of
M31. At $r \sim 250$~arcsec, the existing profiles are simply not reliable
enough to resolve the truncation issue. Whether M32's luminosity profile
is truncated or not therefore remains a wide open question.

$\bullet$ {\sl Since \citet{M 1950}, \citet{S 1954} and
\citet{A 1964}, it has been known that M31's disc exhibits some
rotation curve and optical asymmetries. Although \citet{R 1966},
\citet{RF 1970} and \citet{E 1974} showed these asymmetries to be
much smaller than originally thought, the view that the two
galaxies must be very close neighbours has been reinforced (a) by
recent maps compiled by \citet{G+ 2006} that revealed asymmetries
in M31's dust lanes (as distinct from its spiral arms) and (b) by
computer simulations (\citealt{B 1976} through \citealt{G+ 2006})
that demonstrated that a gravitational encounter between M31 and
M32 could produce these asymmetries .}

This piece of evidence is circumstantial. It does not exclude
alternative explanations for the asymmetries, especially as many
spirals with no major satellite galaxy also exhibit similar
asymmetries in their dust lane distribution (e.g.\ M94) and those
with a large satellite often exhibit asymmetries in the dust lanes
most distant from the satellite (e.g.\ M51).

$\bullet$ {\sl Very recent numerical simulations by
\citet{B+ 2006} suggest that two off-centre inner dust rings in
M31 are likely to be density waves caused by a companion galaxy
colliding head on with the centre of M31's disc 210 million years
ago. M32 is cited as the most likely culprit.}

It should be remembered though that the probability of a head-on
collision having occurred between M31 and M32 is orders of
magnitude lower than that of a chance [approximate] alignment
between the same two galaxies. \citet{B+ 2006} did not rule out
other companion galaxies of M31 as possible culprits and we
therefore believe that it is premature to lay the blame on M32.
For this piece of evidence to be more than circumstantial, a
correlation would need to be established between analogous inner
dust rings in other giant spirals and the presence of disturbed
satellites nearby.

$\bullet$ {\sl M32 does not appear to possess globular clusters
even though it might be expected to possess about 15--20
\citep{H 1991} or 10-20 \citep{Vdb 2000b} such objects [should it
be a dwarf galaxy at the same distance as M31]. This suggests that
M32 must have been tidally stripped of its clusters by M31.}

It should be noted however, that this standard line of argument
was not endorsed by \citet{H 1991} who pointed out that: ``no one
[had] quantitatively demonstrated that any original population of
clusters could have been totally removed by a combination of tidal
stripping from M31 and dynamical friction and tidal shocking from
the dense nucleus of M32 itself''\footnote{Note also that
\citet{HR 1979} estimated that M32 has $\ls$3 globular clusters
(cf.\ the figure of zero normally assumed).}. As this situation is
still the case today, the evidence for the standard argument is
therefore necessarily circumstantial.

If M32 really were three times more distant than M31 though, how
many globular clusters might it be expected to possess?  Since
\citet{H 1977} and \citet{HR 1979} it has been known that the
total number of clusters scales approximately with galaxy
luminosity. It should also be remembered though that ellipticals
in small groups tend to have only about half the number of
clusters of their counterparts in the Fornax and Virgo clusters,
and that some field ellipticals have even fewer \citep{H 1991}. If
M32 were a background galaxy, it would be an isolated system and
might therefore be expected to possess between $\ls$50 and
$\sim$100 clusters.

However, due to distance effects alone, these clusters would be
about 2.5~mag fainter than M31's\footnote{\citet{Vdb 2007} has
found that the brightest metal-poor clusters in NGC~5128 are of
the order of 2~mag intrinsically less luminous than the brightest
metal-rich clusters in the same galaxy. If M32's clusters were
mostly metal-poor, there is then a possibility that the brightest
ones might be of the order of 4.5~mag fainter than M31's brightest
clusters.}. They would therefore only have been relatively
sparsely sampled even by the deepest globular cluster surveys of
the M31 field to date \citep{G+ 2004, K+ 2007}. Also, it is quite
possible that many clusters bright enough to be observed may have
been mistaken for individual stars or blended double or triple
stellar images. Furthermore, especially in the cases of the
faintest definite and candidate globular clusters, surely it is an
assumption that they are all associated with M31, rather than with
M32 in at least some cases? Unfortunately, due to M31 and M32's
close proximity on the sky, their similar recession velocities,
the irregular nature of M31's dust lanes and the serious
reddening/extinction problem, should M32 be a background galaxy
with its own globular cluster system, reliable separation of the
two systems might not be a straightforward task.

It is therefore not known how many (if any) globular clusters M32
has and further study is clearly needed\footnote{Ironically, the
nearby large elliptical galaxy, NGC~5128, was originally thought
to possess few, if any globular clusters \citep{Vdb 1979,H 1991}.
However, following the discovery of its first cluster by
\citet{GP 1980} large numbers of clusters were subsequently
discovered by \citet{VdbHH 1981}, \citet{HHVdbH 1984} and
\citet{HHH 1986}. NGC~5128 is now known to have a large cluster
system which \citet{HHHC 1984} have estimated to comprise of
between 1200 and 1900 clusters.}. Unfortunately, the M31 field is
a problematical one. To quote from \citet{H 1991}: ``With M31, we
encounter a uniquely frustrating morass of observational and
interpretative difficulties involving combinations of sample
contamination, large and uncertain reddening corrections for the
many clusters projected on the M31 disk, and a highly nonuniform
photometric data base.''  With the recent surveys of
\citet{G+ 2004} and \citet{K+ 2007} more uniform databases are
becoming available. However, the uncertainties over reddening and
extinction persist with no immediate prospect of being resolved.

\subsection{Evidence that M32 is in front of M31}
\label{Subsec2.2}

$\bullet$ {\sl Were M32 situated to the rear of M31, it has long
been presumed that it would be intrinsically significantly bluer
than its observed colour, which was measured by \citet{SL 1983} to
be $(B-V)=$0.95~mag. Although a more recent CCD-based study
\citep{P 1993} gives $(U-B)=$0.57~mag (cf.\ the generally accepted
value of 0.50~mag from \citealt{SL 1983}) suggesting that the
actual apparent $(B-V)$ value might be of the order of 0.05~mag
redder than $(B-V)$=0.95~mag, absorption must be small should M32
lie behind M31, as evident from Figure~\ref{f01}(g).
\citet{FJJ 1978} have estimated that M31 would redden M32 by
0.44~mag if M32 lies behind M31. The foreground would contribute
an extra 0.064~mag using an up-to-date value for $E(B-V)$ from
\citet{SFD 1998} implying an intrinsic colour of
0.45$\ls(B-V)\ls$0.50~mag, which is too blue even for an
intermediate elliptical or lenticular galaxy.}

However, if $E(B-V) {\rm (M31)} \sim 0$~mag and
$A_{B}$(M31)$\sim0$~mag, in line with the findings of e.g.\
\citet{X+ 1999} that many spirals are optically thin in their
outer regions, as well as with Figure~\ref{f01}(a), (b) \& (c) and
Figure~\ref{f02}; then M32's intrinsic colour would no longer be
too blue.

$\bullet$ {\sl The lack of absorption clouds seen superimposed
onto M32 has been cited by \citet{FJJ 1978} as evidence that M32
must be in front of M31.}

Most of M31's visible absorption features lie on the side of its
disc opposite to M32 whilst those that lie on the same side are
mainly found at lower radial distances from M31's bulge. Likewise,
the absence of detected optical absorption gradients across M32 is
only circumstantial evidence.

\begin{table}
\caption[]{\baselineskip 3.6mm Published data for planetary nebula
candidates in the direction of M32. Planetary nebulae are ranked
by $m_{5007}$ (from \citealt{CJFN 1989}) when available.
Membership status is generally based on radial velocity
measurements \citep*{NF 1986,HkMH 2004} when available (mean
values having been used when values from both sources are
available). Reddening estimates \citep*{RSMc 1999}, equatorial
coordinates \citep{CJFN 1989,HkMH 2004} and designations, denoted
C for \citet{CJFN 1989} and H-K for \citet{HkMH 2004}, are also
listed.} \label{t1}
$$
{\scriptsize
         \begin{array}{cccccccccl}
            \hline
            \noalign{\smallskip}
$Rank$ & m_{5007} & $Velocity$ & $Member?$ & E(B-V) & $R.A.\ (J2000)$ & $Dec.\ (J2000)$ & $C$ & $H-K$ & $Notes$ \\
       & $(mag)$  & ($km~s$^{-1}) &          & $(mag)$ & ($h~m~s$)    & ($deg~arcmin~arcsec$) & &       & \\
            \noalign{\smallskip}
            \hline
            \noalign{\smallskip}
1  & 20.70 & $-$    & ?     & $-$         & 00~42~41.94 & +40~51~57.9 & 27  & $-$ & $Close to nucleus$ \\
2  & 20.93 & -229   & $Yes$ & 0.01\pm0.01 & 00~42~39.61 & +40~52~39.8 &  5  & $-$ & $-$ \\
3  & 21.04 & -142   & $Yes$ & $-$         &~00~42~40.20 & +40~51~01.9 &  3  &   4 & $-$ \\
4  & 21.09 & $-$    & ?     & $-$         &~00~42~42.24 & +40~51~38.6 & 21  & $-$ & $Close to nucleus$ \\
5  & 21.14 & $-$    & ?     & $-$         &~00~42~55.99 & +40~51~12.1 & 29  & $-$ & $-$ \\
6  & 21.22 & -409   & $No$  & 0.15\pm0.02 &~00~42~31.07 & +40~53~04.5 &  4  & $-$ & $-$ \\
7  & 21.31 & -195   & $Yes$ & 0.17\pm0.03 &~00~42~40.36 & +40~52~43.1 &  6  & $-$ & $-$ \\
8  & 21.39 & -242   & $Yes$ & $-$         &~00~42~42.23 & +40~52~00.7 & 26  & $-$ & $Close to nucleus$ \\
9  & 21.37 & -156   & $Yes$ & 0.31\pm0.10 &~00~42~42.70 & +40~51~56.1 & 25  & $-$ & $Blended with C24$ \\
10 & 21.50 & $-$    & ?     & 0.15\pm0.03 &~00~42~42.54 & +40~51~56.2 & 24  & $-$ & $Blended with C25$ \\
11 & 21.61 & -253   & $Yes$ & $-$         &~00~42~42.25 & +40~51~48.5 & 23  & $-$ & $-$ \\
12 & 22.00 & -193   & $Yes$ & 0.11\pm0.01 &~00~42~35.91 & +40~53~00.1 &  1  & $-$ & $-$ \\
13 & 22.05 & -161   & $Yes$ & $-$         &~00~42~40.09 & +40~49~40.9 &  9  &   3 & $-$ \\
14 & 22.15 & -423   & $No$  & 0.57\pm0.01 &~00~42~26.96 & +40~49~41.5 & 17  & $-$ & $-$ \\
15 & 22.48 & -212   & $Yes$ & $-$         &~00~42~41.68 & +40~51~30.7 & 20  & $-$ & $-$ \\
16 & 22.57 & $-$    & ?     & $-$         &~00~42~23.13 & +40~50~22.2 & 30  & $-$ & $-$ \\
17 & 22.65 & $-$    & ?     & $-$         &~00~42~49.88 & +40~51~10.0 & 22  & $-$ & $-$ \\
18 & 22.83 & $-$    & ?     & $-$         &~00~42~45.00 & +40~52~47.5 & 10  & $-$ & $-$ \\
19 & 22.85 & -154   & $Yes$ & $-$         &~00~42~40.24 & +40~50~59.6 & 14  & $-$ & $-$ \\
20 & 23.28 & -174   & $Yes$ & 0.05\pm0.02 &~00~42~56.99 & +40~51~01.2 & 11  &   7 & $-$ \\
21 & >23.30 & -261  & $Yes$ & $-$         &~00~42~36.50 & +40~51~27.9 & 15  & $-$ & $-$ \\
22 & >23.30 & $-$   & $No?$ & $-$         &~00~42~36.87 & +40~50~02.4 & 19  & $-$ & $\citet{NF 1986} non-member$ \\
23 & >23.30 & -664  & $No$  & $-$         &~00~42~54.58 & +40~52~06.8 & 12  & $-$ & $-$ \\
24 & 23.33  & $-$   & $No?$ & $-$         &~00~42~31.43 & +40~53~22.6 & 18  & $-$ & $\citet{NF 1986} non-member$ \\
$-$ & $-$   & -212  & $Yes$ & 0.06\pm0.01 &~00~42~53.05 & +40~48~59.5 &  2  &   6 & $Not ranked$ \\
$-$ & $-$   & -174  & $Yes$ & 0.03\pm0.01 &~00~42~48.20 & +40~54~12.9 &  7  & $-$ & $Not ranked$ \\
$-$ & $-$   & -151  & $Yes$ & 0.20\pm0.01 & 00~43~02.10 & +40~49~30.4 &  8  &   8 & $Not ranked$ \\
$-$ & $-$   & -181  & $Yes$ & $-$         &~00~42~44.51 & +40~48~07.7 & 13  & $-$ & $Not ranked$ \\
$-$ & $-$   & -182  & $Yes$ & $-$         &~00~42~23.24 & +40~46~28.7 & $-$ &   1 & $Not ranked$ \\
$-$ & $-$   & -158  & $Yes$ & $-$         &~00~42~28.93 & +40~44~39.0 & $-$ &   2 & $Not ranked$ \\
$-$ & $-$   & -184  & $Yes$ & $-$         &~00~42~50.63 & +40~45~28.4 & $-$ &   5 & $Not ranked$ \\
            \noalign{\smallskip}
            \hline
     \end{array}
}
     $$
   \end{table}

$\bullet$ {\sl Some planetary nebula candidates detected in the
M32 field are not sufficiently reddened for them to lie behind M31
(based on the estimate of \citealt{FJJ 1978} that M31 would redden
objects behind it by 0.44~mag).}

As this piece of evidence is potentially conclusive, it deserves
careful attention. The published data available for planetary
nebulae in the M32 direction prior to those of \citet{M+ 2006} are
presented in Table~\ref{t1}. On the one hand, the very low
$E(B-V)$ value of $0.01(\pm0.01)$~mag derived for the 2nd ranked
object (an almost certain member of M32) and the much higher
values derived for two definite non-members, support the
prevailing view that M32 must lie in front of M31. However, the
corresponding values of $0.17(\pm0.03)$, $0.31(\pm0.10)$,
$0.11(\pm0.01)$ and $0.20(\pm0.01)$~mag, derived for the 7th, 9th
and 12th ranked objects and an unranked object respectively, all
four of which are almost certainly members, are significantly
larger than the Galactic reddening component of 0.064~mag (from
\citealt{SFD 1998}) and require 0.11, 0.25, 0.05 and 0.14~mag
respectively of additional reddening. This extra reddening
component is unlikely to be due to M32's interstellar medium as
none has been detected to date \citep*{E 1974,SWM 1998} and so if
M32 is in front of M31 then it must be intrinsic to the planetary
nebulae. Clearly, there are some inconsistencies with the existing
picture that need further investigation. Could M31's disc be
optically thin in places and/or could some of the objects be
unresolved H~II regions in M32 or (in the absence of velocity
data) even in M31? Certainly, from a recent ultra-violet
photometric study by \citet{Gdp+ 2005} we would expect a large
number of M31's H~II regions to be projected onto M32.

\section{Evidence based on distance indicators}
\label{Sec3}

\subsection{Planetary nebulae}
\label{Subsec3.1}

$\bullet$ {\sl M32's planetary nebula luminosity function (PNLF)
has been used by \citet{CJFN 1989} to investigate approximate
limits on M32's distance. The 1-$\sigma$ maximum-likelihood
confidence contours presented in their figure~6 corresponded to
distances of $0.91_{-0.30}^{+0.10}$~Mpc for M32 and
$0.72_{-0.04}^{+0.02}$~Mpc for the bulge of M31.}

A more accurate estimate of $m^{*}_{5007}$ was not possible for
M32 because of the small planetary nebula sample size. However,
the 1-$\sigma$ error bars would have been even larger for M32 if
membership had not been assumed for objects lacking velocities
e.g.\ the 1st, 4th, 5th and 10th ranked objects in Table~\ref{t1}.
Also, based on the means of the $E(B-V)$ values derived for
subsets of both the M32 and M31 bulge planetary nebula samples
\citep{RSMc 1999} the M31-bulge objects typically suffer from
0.2~mag {\sl more\/} extinction at $\lambda=5007$~\AA\ than do the
M32 objects. If this were taken into account, M32's PNLF-based
distance relative to that of M31's bulge would increase by 10~per
cent. In the absence of a reliable value of $m^{*}_{5007}$, it may
still be a useful exercise to compare M32's brightest definite
member planetary nebula, for which $m_{5007} = 20.93$~mag and
$E(B-V) = 0.01 \pm 0.01$~mag, with its counterpart in the bulge of
M31, for which $m_{5007} = 20.28$~mag and $E(B-V) = 0.20 \pm
0.02$~mag \citep{CJFN 1989,RSMc 1999}. This leads to a distance
ratio of about 1.8 for M32's distance with respect to M31's.
Although using the 1st ranked object in Table~1 (whose membership
of M32 has yet to be confirmed) would reduce this distance ratio
by 10 per cent, using a lower ranked M31 planetary nebula with a
higher $E(B-V)$ value, e.g.\ M31~PN~31 for which
$m_{5007}=20.49$~mag but $E(B-V)=0.40 \pm 0.25$~mag
\citep{CJFN 1989,RSMc 1999} would increase the ratio by about
30~per cent. Bearing in mind that $E(B-V)$ values are only
available for a small fraction of M31-bulge planetary nebulae,
this 30~per cent figure should be treated as a lower limit. This
suggests that if M32 really lies at a distance of about 0.76~Mpc
then its PNLF is intrinsically $\gs$1.5~mag fainter than that of
M31's bulge. Alternatively though, both galaxies' PNLFs could be
similar if M31 is optically thin in the M32 field and M32 lies
well to the rear of M31 at a distance of the order of
$\gs$1.5~Mpc.

Very recently, \citet{M+ 2006} have published $m_{5007}$ measurements
for 46 emission-line objects that are probable members of M32 based
on their projected positions and measured radial velocities. Of these,
4 are extended objects, leaving 42 planetary nebula candidates.
The brightest planetary nebula candidate is of $m_{5007}=20.78$~mag
(cf.\ 20.70~mag for the first ranked object in Table~\ref{t1})
whilst the majority of the candidates (31 out of 42) are of 23rd,
24th, 25th or 26th~mag. This new dataset therefore confirms our
finding that planetary nebulae in M32 appear to be systematically
fainter than their counterparts in the bulge of M31, consistent with
M32 being at least of the order of twice as distant.

\subsection{RR Lyraes}
\label{Subsec3.2}

$\bullet$ {\sl RR Lyraes detected in the M32 direction by
\citet*{AgMW 2004} were found to be of very similar apparent
magnitudes to others detected in an M31-only field. If RR Lyraes
have indeed been found in M32 as presumed, then there would be a
very strong case for M32 and M31 being at similar distances. This
study was based on HST observations.}

The main issue here is over the membership status of the 22 RR
Lyraes found in the M32$\cup$M31 field cf. the 10 found in the
M31-only control field, or in other  words, whether this
over-density is necessarily due to M32. We note that if M32's
distance were three times that of M31, its RR Lyraes would be too
faint to have been detected. All of the RR Lyraes found would
therefore have to be members of M31. Such a scenario cannot be
ruled out because we would expect large differences in the column
number densities of detectable RR Lyraes in different M31-only
fields due not only to Poisson statistics but also to differences
in absorption. Stronger absorption would reduce the detectable
number of RR Lyraes, particularly those on the far side of M31. It
may therefore be significant that the control sample was taken
from a field affected by visible absorption regions whilst the M32
sample was taken from a field devoid of absorption features (see
fig.~1 of \citealt{AgMW 2004}). Further evidence for this
difference in absorption includes the greater redward spread in
$(V-I)$ exhibited by stars in the control field (see fig.~2 of
\citealt{AgMW 2004}).

\subsection{Surface-brightness fluctuations}
\label{Subsec3.3}

$\bullet$ {\sl A distance of 0.83~Mpc was derived by
\citet{TS 1988} for M32 based on surface-brightness fluctuations
(SBFs) observed in CCD images of the core of M32 with its nucleus
excised. This estimate probably constitutes the first formal
distance measurement for M32. Consistent results were also found
by \citet{T 1991} and \citet{LT 1993} using $I$ and $K$-band
observations respectively.}

Any SBF distance determination requires a prior estimate of the bright
end of the stellar luminosity function (LF) in the target galaxy concerned.
In practice, this information is generally based on the known stellar LFs
of resolved Local Group galaxies of the same morphological type and colour,
and whose distances are known (based on independently made distance measurements).

At the time of the first SBF distance determinations for M32,
red-giant branch (RGB) stars had yet to be resolved in any [known]
intermediate/giant early-type galaxy and had only just been
resolved in M32 [whose type and distance we necessarily treat as
unknown here] by \citet{F 1989} and \citet{DJ 1992} using the
Canada-France-Hawaii telescope (CFHT). Also, M32's intrinsic
colour and the value of $A_B$(M32) were of course unknown as they
still are today (see Section~\ref{Subsec2.2}). The required LF
information was therefore unavailable. Note that in order to
calibrate the absolute magnitude scale of M32's stellar LF from
stellar photometry, one would need to assume M32's distance a
priori--a procedure that would clearly invalidate any SBF distance
determination based on such an LF. Consequently, assumptions about
the bright end of M32's stellar LF had to be inferred from
galaxies of other morphological types.  The SBF distance
determinations of \citet{TS 1988}, \citet{T 1991} and
\citet{LT 1993} therefore awaited confirmation (1) that the
morphology of M32's colour-magnitude diagram (CMD) was similar to
that previously found in other Local Group galaxies (of different
morphological types) including M31, as opposed to objects with
anomalous SBFs suggestive of much brighter stellar populations,
such as the isolated RCG, NGC~4489
\citep{PM 1994,JLT 1996,JTL 1998,MSQ 2001}; and (2) that the CMD
morphologies of intermediate ellipticals would turn out to be
similar to that of M32.

Confirmation that these assumptions appear to have been well
justified came in two stages. First, the HST stellar photometry of
\citet{G+ 1996}; \citet{AgMW 2004} and \citet{WMAgE 2004} found
M32's CMD morphology and stellar apparent LF to be very similar to
that of M31. More recently, \citet{Rej+ 2005} have resolved RGB
stars in the halo of the nearby peculiar intermediate
elliptical/lenticular NGC~5128 using the HST. Their study found an
upper-RGB CMD morphology similar to that of M31 and M32. This
finding is quite significant although it has two possible
limitations. First, NGC~5128 is an active galaxy--the giant double
radio source Centaurus~A, which is believed to be a merger between
an intermediate elliptical and a spiral galaxy. It is therefore
certainly not an ideal model galaxy, but unfortunately there are
no better candidates as no normal intermediate/giant elliptical
galaxy is near enough for resolved stellar photometry. Secondly,
the field studied was 33.3~arcmin (or $\sim$38~kpc) from the
centre of NGC~5128. \citet{Vdb 1976} found the effective
radius ($r_e$) of NGC~5128 to be 5.5~arcmin, so the halo field
corresponds to a radial distance of about 6$r_e$. This would
correspond to a field at about 3~arcmin from the centre of M32,
whereas the SBF distance measurements were based on the core of
M32. If there are often major differences between the stellar
populations of the cores of intermediate/giant ellipticals and
their haloes, then the upper RGB stars observed in NGC~5128 may
not necessarily be representative of the brightest stellar
populations in the core of  M32. Evidence has already been found
for variations in stellar chemical composition and mean age (hence
stellar LF too) between the inner and outer regions of M32
\citep{DDrY 1990, D 1991, DJ 1992, DN 1992, Ros+ 2005} though the
variations found are not large enough to invalidate the finding
that M32 is at a similar distance to M31.

\subsection{Red-giant-branch stars}
\label{Subsec3.4}

$\bullet$ {\sl Using the 2.5-m Mount Wilson telescope, \citet{B 1944}
obtained red-sensitive photographic plates of M32, NGC~205 and the central
region of M31 itself. On these plates, he found large numbers of faint
stellar images associated with each galaxy. The images were just above
the detection limit of his red plates but were not present on blue-sensitive
plates of the same
fields taken previously. This finding suggested that M32, NGC~205
and M31 had all been resolved into red-giant stars of similar
magnitude ($m_{pg}\sim$21.3~mag), indicating that they were all at
similar distances.}

In the 1940s, red giants were considered to be a homogeneous type
of star and therefore any red giant was considered to be a viable
distance indicator. We now know that there is a considerable range
in intrinsic luminosity (and even colour) for red giants,
depending on what the masses and abundances of their progenitors
were and what stage of evolution they are in. A further
complication might be variations in stellar chemical composition
and mean age (hence stellar LF too) between M32, M31 and NGC~205;
as well as between the inner and outer regions of M32 itself
\citep{DDrY 1990,D 1991,DJ 1992,DN 1992,Ros+ 2005}. Therefore, the
brightest red giants detected cannot be used as distance
indicators in the straightforward way that \citet{B 1944} used
them. At the very least, CMDs and/or stellar LFs are needed.

In the case of M32 (as opposed to NGC~205) crowding and contamination by
stars in M31 are also major issues. \citet{B 1944} noted that: ``the central
part of M32 is completely burnt out'', i.e.\ no stars could be resolved in the
region where stars from M32 heavily outnumber those from M31. He further noted
that: ``at greater distances from the center of M32 the members of the two
systems are hopelessly mixed''. For intermediate radial distances, we note
that crowding would affect annular regions centered on M32, which would appear
to be centered on the centroid of M32--even if a significant component were due
to stellar images from M31 merged with others due to M32. The brightest point-like
images could therefore have been artifacts of both crowding and contamination by M31.

In the absence of reproductions of the plates taken by
\citet{B 1944} and information detailing specifically which fields
were used to identify the red-giant members of M32, it is not
possible to quantify the degree of crowding or contamination by
M31. However, it is probably significant that in the optical, red
giants were not resolved unambiguously in M32 [again] for another
fifty years--until HST images became available in the mid 1990s.
Although successful in detecting bright RGB stars in M32 and M31,
even the CFHT 3.5-m {\it VRI}~stellar photometry of
\citet{F 1989} was subsequently found to suffer from the effects
of crowding \citep{G+ 1996}\footnote{By contrast, ground-based
infra-red stellar photometry of M32 \citep{F 1992,DN 1992,ES 1992}
has generally been more successful as it does not appear to have
suffered from significant crowding.}.

$\bullet$ {\sl A distance of 0.63$\pm$0.6~Mpc (with minor
adjustments for different assumed metallicities and different
interpretations of the data) was derived for M32 by \citet{F 1989}
based on [ground-based] CFHT {\it VRI}-band stellar photometry of
RGB stars in M32 and M31. In a major advance on the earlier study
of \citet{B 1944}, RGB stellar LFs were used instead of the vague
concept of the `brightest' stars.}

A clear and concise discussion of those issues in \citet{F 1989}
relevant to the distance determination is given by \citet{F 1990}
to which readers are referred. As explained by \citet{F 1990} in
order to proceed with the distance estimation, an assumption was
needed regarding the type of stellar population being observed. On
the assumption that the tip of the RGB was being observed, two
slightly different alternative distance estimates were obtained,
based on the interpretation of two separate minor discontinuities
found in the derived LFs (at $I=20.1$ and 20.5~mag) as the tip of
the RGB.

Although the ground-based photometry of \citet{F 1989} was found
by the later HST-based study of \citet{G+ 1996} to suffer from
some crowding problems, the basic assumption of \citet{F 1989}
that the tip of the RGB had been observed was confirmed by
\citet{G+ 1996}. Whilst the ground-based photometry only
penetrated 1-to-2~mag below the observed discontinuities, the
photometry of \citet{G+ 1996} was about 3-to-4 mag deeper.

$\bullet$ {\sl HST Wide-field Planetary Camera 2 (WFPC2) optical
stellar photometry of M32 and M31 has yielded stellar LFs that are
so similar that no difference in [relative] distance was
detectable \citep{WMAgE 2004}. This result is supported by CMDs
also constructed from WFPC2 photometry which show little
difference in morphology or apparent magnitude between the RGB
stars resolved in M32 and M31 \citep{G+ 1996,AgMW 2004,WMAgE 2004}.
Deeper stellar photometry obtained very recently using the HST's
Advanced Camera for Surveys (ACS) also supports this result (Worthey,
private communication).}

Both WFPC2-based studies (\citealt{G+ 1996} on the one hand and
\citealt{AgMW 2004} and \citealt{WMAgE 2004} on the other) derive
$(V-I)$ versus $I$ CMDs for M32 and M31, yielding a single RGB of
similar apparent brightness for each galaxy. Now, let us consider
what kind of CMDs we might expect to see should M32 lie at three
times the distance of M31. Assuming that the RGBs in both galaxies
are of similar intrinsic brightness\footnote{Until very recently,
it was impossible to predict the precise nature of the stellar
populations in normal intermediate or giant early-type galaxies
because none had been resolved. However, as mentioned in
Section~\ref {Subsec3.3}, RGB stars have very recently been
resolved in the halo of the nearby galaxy, NGC~5128 by
\citet{Rej+ 2005}. The results suggest that its upper-RGB CMD
morphology is very similar to that already found for M31 and M32
by \citet{G+ 1996}, \citet{AgMW 2004} and \citet{WMAgE 2004}.
Although NGC~5128 is by no means an ideal model galaxy, the
absence of any brighter hitherto unknown red-giant population
suggests that intermediate and giant ellipticals probably
possess similar RGBs to those of M32 and M31. However, we note
that early-type galaxies with anomalously bright, but as yet
unresolved, populations seem to exist, in the form of the RCG,
NGC~4489 \citep {PM 1994, JLT 1996, JTL 1998, MSQ 2001}.}, M32's
RGB would be observed to be about 2.5~mag fainter than M31's,
making it largely undetectable. This is because for a star to
appear on a $(V-I)$ versus $I$ CMD, not only must it be bright
enough in $I$, but it must also be detected in $V$ (in order to
have a colour assigned to it at all).  Figure~8 of \citet{G+ 1996}
incorporates a $V=26$~mag approximate cutoff, shown as a diagonal
dotted line. If the RGB found for M32 (or that found for M31) were
shifted 2.5~mag faintward, it would lie completely below this
dotted line. Similar selection effects must apply to figure~2 of
\citet{AgMW 2004} (which shares the same dataset as
\citealt{WMAgE 2004}). This would mean that the RGBs observed in
the CMDs for the M32$\cup$M31 field with an estimate for M31's
component subtracted, could not be comprised of stars from M32.
They would have to be comprised of crowding peaks and/or members
of M31. Is such a scenario possible though?

Crowding that was not taken account of by the stellar photometry
software would result in the merging of fainter stellar images
into brighter peaks. This would affect the denser M32$\cup$M31
fields more than the M31-only control fields, and M32's RGB would
appear brighter than it really was. However, in order to achieve a
brightening of 2.5~mag, typically 10 red giants would be needed
per crowding peak. Such a level of crowding would presumably alter
the morphology of the RGB by stretching it in magnitude space. As
the observed RGBs do not show obvious signs of broadening in
magnitude space, crowding is probably not a serious problem.

On the other hand, contamination due to M31 could be quite
significant, because each WFPC2-based study was based on a single
M31-only control field. Also, in each case, the correction for
contamination by M31 was derived from the M31-only control field
by the application of a simple multiplicative factor designed to
account for the differences in the surface-brightness of M31 based
on elliptical-isophote surface photometry. As $A_{B}$(M31) was an
unknown quantity throughout, no account could be taken of the
irregularities in M31's surface brightness or of the differences
in absorption between the control fields and the M32$\cup$M31
fields. As already mentioned in Section~\ref{Subsec2.2}, the M32
direction appears to be devoid of absorption features and this may
account for part of the over-density of bright resolved objects
found in the M32$\cup$M31 fields with respect to the M31 fields.
Note also that the M32$\cup$M31 fields were necessarily located
far away enough from the centroid of M32 for severe crowding to be
avoided. Whilst the M32$\cup$M31 field of \citet{G+ 1996} spanned
a range in radial distance from M32's centre of about 1 to 2
arcmin, the field of \citet{AgMW 2004} and \citet{WMAgE 2004} was
further out and spanned a range of about 3 to 4 arcmin.
Particularly in the latter case, contamination due to M31 could be
extremely serious. \citet{AgMW 2004} estimated that 41\% of the
stars found in their M32 $\cup$M31 field were due to M31, but,
this figure is understandably highly uncertain. Whilst a
significant upward revision of this percentage would probably not
affect significantly the morphology of the inferred red clump for
M32 alone, it could cause significant thinning of the much more
sparsely populated RGB. However, it is very unlikely that this
thinning could be sufficient to shift the tip of the RGB faintward
by as much as 2.5~mag (albeit $<$2.5~mag, should the M31-only
control field suffer from significant absorption).

We therefore find that although the WFPC2-based studies had some
limitations, their findings nevertheless provided the first
corroborative evidence in support of the existing view that M32
and M31 are at similar distances. Also, in a new development, we
understand that recent ACS observations that reach even deeper
than $I \sim 28$~mag do not show any sign of a second, fainter,
RGB, which would be expected if M32 were a more distant galaxy
(Worthey, private communication). When published, this new ACS
photometry should provide further corroborative evidence that
M32 is not a background galaxy.

\section{Red compact dwarfs beyond the Local Group}
\label{Sec4}

$\bullet$ {\sl The existence of five other RCD candidates beyond
the Local Group; namely NGC~4486B, NGC~5846A, two objects in
Abell~1689 discovered by \citet{M+ 2005} and one object in
Abell~496 discovered by \citet{C+ 2007}\footnote{Note that giant
or intermediate RCGs are too large to be relevant to the present
discussion. Likewise, ultra-compact dwarfs (UCDs) are too small.
UCDs, which were discovered in the Fornax cluster by
\citet{H+ 1999} and \citet{DJGP 2000} are typically 5~mag less
luminous than M31's main companions.}; suggests that M32's
compactness is not unique and that its defiance of most known
scaling relationships defined by normal early-type galaxies is
therefore not a problem.}

This line of argument assumes that it has been established that
the other candidates are definitely RCDs. Whilst it would be
strengthened if it were proven that these other objects were
indeed genuine RCDs, it would not follow that M32 must also be
one--especially as M32 would have to be the most extreme example
of all (cf.\ fig.s~3 and 4 of \citealt{C+ 2007}). Likewise, if it
were ever demonstrated that M32 were definitely a background
galaxy, it would not follow that other RCD candidates must also be
background galaxies. Linkages of this nature can at best only
provide circumstantial evidence.

It could also be argued that the extreme rarity of RCD candidates
calls into question the very existence of this morphological type
of galaxy. In fact \citet{G 2002} has already called into question
the existence of compact ellipticals (cf.\ lenticulars).
Considering the vast numbers of galaxies that have been observed
to date (including a large number of ongoing and recent mergers)
and that there have been [unsuccessful] systematic searches for
RCDs in nearby clusters \citep{DG 1998,ZB 1998} the scarcity of
examples may be more significant than the existence of several RCD
candidates.

We note that whilst the extreme rarity of RCD candidates could be
due to abnormal physical conditions that only last for relatively
short periods, e.g.\ galaxies in the process of being tidally
stripped by larger neighbours\footnote{Clear evidence for tidal
truncation has yet to be demonstrated for any RCD candidate
including M32. Although \citet{KK 1973} found the profile of
NGC~5846A to be significantly truncated, this result may have been
due to sky-level determination problems \citep{CDv 1983,PNS 1987}.
Also, \citet{PNS 1987} found that the profile of  NGC~4486B did
not appear to be severely truncated, though they found that the
profile of an RCG also in the vicinity of M87 (i.e.\ NGC~4486),
namely NGC~4478, did appear to be severely truncated. This latter
galaxy is intrinsically about 2~mag brighter than NGC~4486B
\citep{PNS 1987} (and about 3~mag brighter than M32 if it is at
the same distance as M31). It is therefore an intermediate
galaxy.}, it could possibly be the product of some rare
observational selection effects, e.g.\ galaxies' discs being
submerged beneath the signal and noise of the bright discs or
haloes of larger neighbours and the inherent difficulties involved
in subtracting off the `background' light from these giant
neighbours. We would like to see a definitive study of the
`background'-subtraction issues in order to rule out (or rule in)
this possibility. Note also that the fact that there are so few
RCD candidates allows for a different explanation for each
individual case.

$\bullet$ {\sl Based on certain well-known scaling relationships,
RCDs (as opposed to normal [lower surface-brightness] dwarf
ellipticals) are known to be the low-luminosity relatives of
normal giant ellipticals \citep{WG 1984,K 1985}. RCDs, including 
M32, are therefore not abnormal at all as they are structurally 
similar to giant ellipticals.}

This issue has already been dealt with briefly in
Appendix~\ref{AppA} and readers are referred to \citet{GG 2003}
and \citet{G 2005}. To this it should be added that the notion of
a dichotomy between normal early-type giants on the one hand,
versus normal [low surface-brightness] early-type dwarfs on the
other, has been renounced even by its staunchest proponents (see
e.g.\ \citealt{JB 1997} cf.\ \citealt{BC 1991}).

\section{Summary, discussion and conclusions}
\label{Sec5}

Prior to \citet{B 1944} it was not known whether M32 was a dwarf
satellite of M31 or a background galaxy. However, following
\citet{B 1944} it became established as an unquestionable fact
that M32 must be a dwarf satellite of M31. We have demonstrated
that contrary to popular belief, the only robust evidence for this
case is very recent in nature and ultimately rests entirely
on resolved stellar photometry obtained with the HST. In other
words, in the absence of these HST observations, the hypothesis
that M32 might be a background galaxy would still be irrefutable.
The assumptions on which the distance measurements derived from
SBFs \citep{TS 1988,T 1991,LT 1993} and ground-based resolved
stellar photometry \citep{F 1989} were based, have been confirmed
by later HST observations\footnote{As SBF studies amount to
unresolved stellar photometry, they are related to (and
complementary to) resolved stellar photometry but do not
constitute independent evidence.}, but all of the preceding
evidence beginning with that of \citet{B 1944} was found to be
circumstantial.

We accept that M32 could well be an RCD satellite of M31.
Unfortunately however, we would be left with the well known
problem of explaining why M32 (and to a lesser extent five other
less extreme RCD candidates) should defy most known scaling
relationships defined by normal early-type galaxies. This is
regardless of whether M32 is an elliptical or a lenticular.
Presumably this might be because RCDs are in the process of being
tidally stripped and are therefore stellar systems that are not in
equilibrium. More specifically though, a major problem we are left
with is why as a dwarf galaxy, M32 should have a central stellar
velocity dispersion several times higher than that of M31's other
main companions (which are of similar apparent brightness to M32).
Could this be because M32 used to be a much larger normal galaxy
prior to the commencement of the stripping process? If so, perhaps
it could have retained its original central velocity dispersion.
However, such a scenario would contradict the findings of
\citet{NP 1987} and \citet{CGJ 2002} who concluded that the
precursor of  M32 was unlikely to have been a [larger] normal
intermediate/giant elliptical galaxy. Alternatively,
\citet{B+ 2001} have suggested that the precursor of M32 might
have been a late-type spiral bulge. If so, this might help to
explain M32's anomalously high central velocity dispersion.
Another outstanding issue though is why M32's planetary nebulae
appear to be systematically fainter in apparent magnitude than
M31's if the two galaxies are at the same distance.

The main arguments in favour of M32 being a normal background
galaxy are as follows.

(1) M32's central velocity dispersion is typical of intermediate
galaxies (including  the bulges of lenticulars), and not of
dwarfs. The homogenized mean value listed by the Hyperleda
database is 72~km~s$^{-1}$ cf.\ 23~km~s$^{-1}$ for NGC~205,
20~km~s$^{-1}$ for NGC~185 and 22~km~s$^{-1}$ for NGC~147
(homogenized mean values from Hyperleda)\footnote{The Lyon-Meudon
Extragalactic Database (Hyperleda) is maintained by the Centre de
Recherches Astronomiques de Lyon and available online at
http://leda.univ-lyon1.fr}. However, recent high-resolution
measurements have yielded even higher values for M32 e.g.\ 126
$\pm$10~km~s$^{-1}$ \citep{Vdm+ 1997} and 130~km~s$^{-1}$ (based
on a Gaussian fit) or $\gs$175~km~s$^{-1}$ (when corrected for the
wings) \citep{J+ 2001}.

(2) Published planetary nebula observations from a variety of sources
\citep{CJFN 1989,HkMH 2004,M+ 2006} are all consistent in suggesting 
that M32 is at least twice as distant as M31.

(3) M32 would obey known scaling relationships defined by normal 
intermediate/giant early-type galaxies including the Fundamental plane 
\citep{DD 1987} if it were of the order of three times more distant than M31.

(4) A physical explanation for M32's apparent compactness would no longer be
needed.

However, this scenario cannot explain the results of space-based resolved stellar
photometry of M32, which find its RGB to be very similar to M31's in both general
morphology and apparent brightness--thereby requiring that the two galaxies are
at almost identical distances. Only in the event of these results becoming
open to re-interpretation could the new hypothesis become a viable explanation.

Unfortunately, as is evident from Figures~\ref{f01}, \ref{f02} and
\ref{f03}, there does not appear to be any middle ground between
these two scenarios. If M32 were behind M31 but close enough to
M31 to be interacting with it, explanations for M32's high central
velocity dispersion and faint planetary nebulae would still be
needed. We therefore find that we are left with two apparently
irreconcilable scenarios, only one of which can be correct, but
both of which suffer from potentially fatal evidence to the
contrary.  This suggests that current understanding of some
relevant fields is inadequate and that further study is urgently
needed.

As far as further work on M32 is concerned, the optical
thicknesses of the relevant regions of M31 are critical quantities
that remain to be determined, and more studies on this problem
would therefore be welcomed. Another remaining long-standing
problem with wide-ranging implications is how the irregular
component of the light contribution from M31 should be accounted
for. The most rigorous subtraction of light from M31 to date is
probably that performed for the surface photometry of
\citet{CGJ 2002}\footnote{Following the example of
\citet{CGJ 2002} meaningful tests of the viability of models for
M31 and M32's light distributions are possible. Synthetic images
can be generated for the entire M31$\cup$M32 field, in which (1)
only M32 has been subtracted, (2) only M31 has been subtracted,
and (3) both M32 and M31 have been subtracted.}. As these authors
noted though, their elliptical isophotal model of M31 was unable
to account for fine-scale structure such as dust lanes and spiral
structure. With this in mind and on account of the newly-found
importance of the HST-based stellar photometry, there is clearly a
need for future space-based resolved stellar population studies,
to obtain  not just one, but several such M31-only control fields
and at more carefully chosen locations. Longer telescope-time
allocations will therefore be needed.

\section{Acknowledgements and author credits}

We thank Guy Worthey, Alister Graham and the referee for useful 
comments, as well as the former director of the NAOC, Guoxiang Ai, 
for funding this project. The review part of this paper was the 
work of CKY and MJC with valuable input from RJD. CKY, MJC, ALL 
and TJZ all contributed to Appendix~\ref{AppA}.

\appendix

\section{Distance estimation based on scaling relationships}
\label{AppA}

\begin{figure}
\centering
\includegraphics[height=110mm,angle=-90]{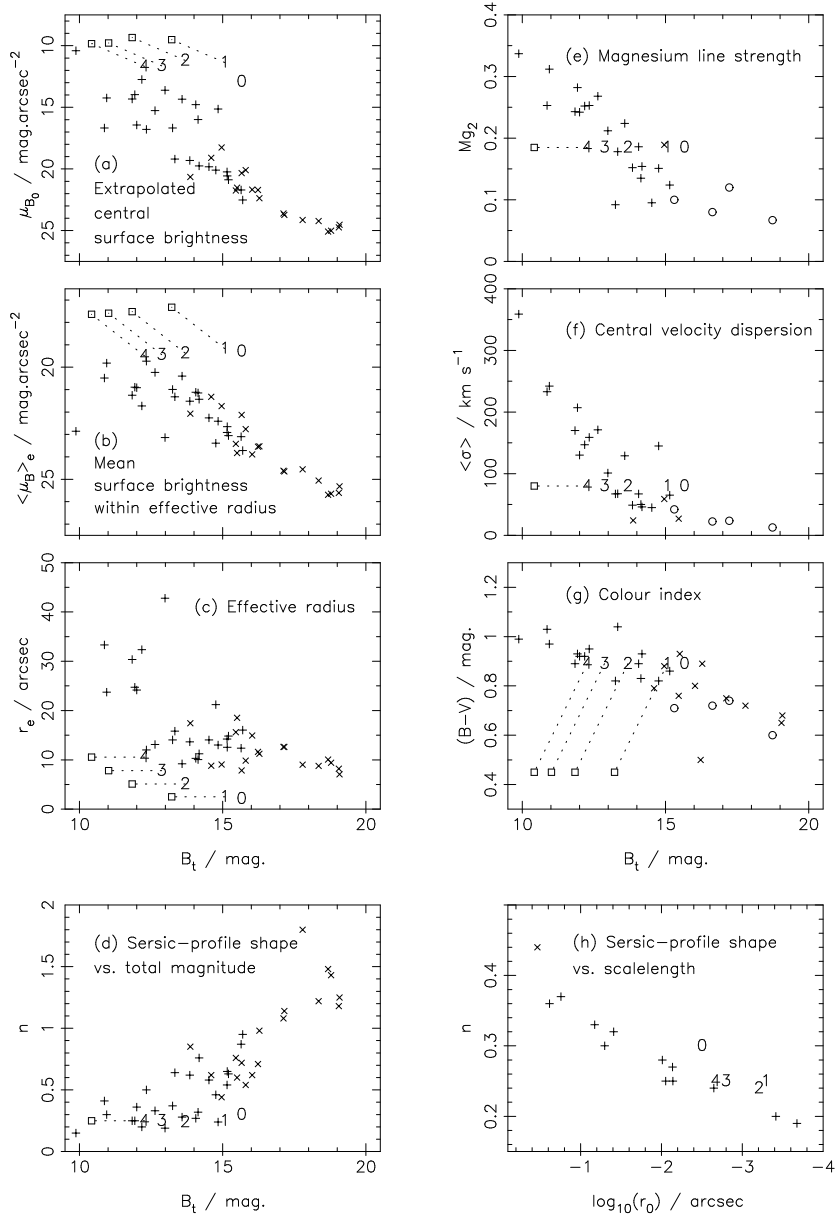}
\vspace{-4mm}

\caption{\baselineskip 3.5mm Expected positions (after relocation
to the FC's distance of 20~Mpc) of M32 as a function of its true
distance (which is unknown) relative to two-parameter
relationships for 43 early-type FC galaxies (including Es, dEs,
S0s and dS0s). Numeral symbols `$0$' denote a true distance of
0.76~Mpc whilst the other integer symbols denote integral true
distances in Mpc, all assuming $A_B$(M31)=0~mag; `$\Box$' symbols
denote corresponding distances of 1 through 4~Mpc [or 4~Mpc only
in the cases of (d), (e), (f) and (h)] assuming
$A_B$(M31)=1.85~mag from \cite{FJJ 1978}; whilst the dotted lines
denote $0.0<A_{B}$(M31)$<1.85$~mag. `$+$' and `$\times$' symbols
denote \cite*{CCDo 1994} and \cite{CB 1987} $B$-band FC
surface-photometry data respectively; the former, whose sky
subtraction procedures were superior, having been used for the
nine common objects. `$\circ$' symbols denote the Local Group
galaxies NGC~205, 185 and 147, as well as the Fornax dwarf
spheroidal, when published data were available. Magnesium line
strengths, `Mg$_{2}$'; central velocity dispersions, `$\langle
\sigma \rangle$' and `$(B-V)$' colour indices were collated from
Hyperleda. For M32, the Hyperleda values of $\langle\sigma\rangle$
exhibited a large scatter, so we adopted the median value of
81~km~s$^{-1}$. `$B_t$', `$\mu_0$', $r_0$ and `$n$' denote the
\cite{S 1968} parameters total magnitude, central surface
brightness, scale length and profile shape respectively; whilst
`$r_e$' and `$\langle\mu\rangle_e$' denote effective radius and
mean surface-brightness within $r_e$ respectively.} \label{f01}
\end{figure}

First, in Figure~\ref{f01}, we consider a wide range of known
two-parameter relationships for early-type galaxies based on a
sample of 43 Fornax Cluster (FC) elliptical and lenticular
galaxies. This sample includes giants, intermediates and [normal]
dwarfs; which are currently accepted as defining a continuous
sequence in S\'ersic luminosity-profile shape index $n$
\citep{S 1968} as well as in many other physical properties, as
discussed by \citet{GG 2003} and \citet{G 2005}\footnote{According
to the traditional view \citep{WG 1984,K 1985} that red compact
dwarfs such as M32 (as opposed to normal dwarf ellipticals) are
the low-luminosity relatives of giant ellipticals, M32 should obey
most of the scaling relationships defined by giant ellipticals
whereas normal dwarf ellipticals should defy them. However, as
evident from Fig.~\ref{f01}, this is clearly not the case.}.
Recent measurements place the FC at a distance of about 20~Mpc
\citep{F+ 2001,T+ 2001,J 2003}. In this section, we investigate
where M32 would lie with respect to these relationships, if it
were re-located to a distance of  20~Mpc and imaged at the same
resolution as the FC galaxy observations used. However, instead of
assuming the actual distance for M32 to be 0.76~Mpc, we have
treated this quantity as an unknown variable lying somewhere
within the range 0.76 to 4.0~Mpc. We find that M32 defies all of
the relationships except for Figure~\ref{f01}(g), the
colour-magnitude diagram (CMD) (which it fits only very poorly) if
its actual distance were about 0.76~Mpc, but would fit all of them
including the CMD (which it would fit extremely well) should its
real distance be about 2-to-3~Mpc and $A_{B}$(M31) be small.
Although most of the two-parameter correlations are not strong
enough for the purpose of  accurate distance estimation, several
of them are nevertheless of great relevance because they have long
been used (e.g.\ by \citealt{BC 1991}) to demonstrate and
characterize the anomalous characteristics of RCGs.

\begin{figure}

\vs\vs \centering
\includegraphics[height=85mm,angle=-90]{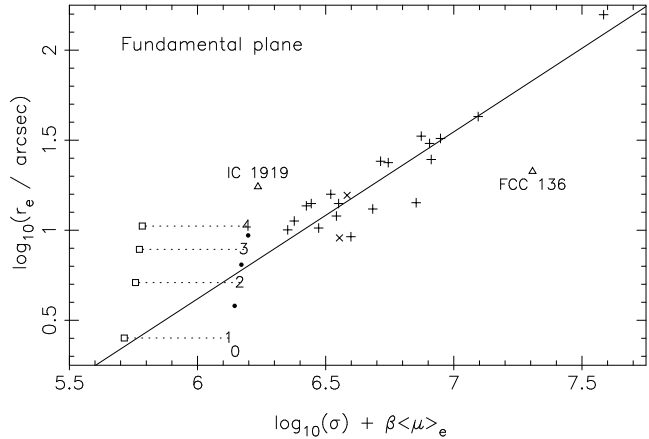}
\caption{\baselineskip 3.6mm Expected positions (after relocation
to the FC's distance of 20~Mpc) of M32 as a function of its true
distance (which is unknown) relative to the Fundamental plane,
$\log_{10}r_{e}=0.928(\log_{10} \langle \sigma \rangle +0.220
\langle \mu \rangle_{e})-4.95$, for early-type FC galaxies (both
Es and S0s) with velocity dispersion measurements. The notation
follows Fig.~\ref{f01} except that the `$\bullet$' symbols denote
true distances of 1.5, 2.5 and 3.5~Mpc; all assuming
$A_B$(M31)=0~mag. The solid line represents the best $r_e$-space
minimized least-squares planar fit to the FC galaxies after the
removal of the two outliers (`{\tiny$\triangle$}' symbols).}
\label{f02}
\end{figure}

Applying the three-parameter Fundamental plane of \citet{DD 1987}
leads to an absorption-dependent distance estimate of
$2.3-0.77(A_{B}$(M31)~mag$^{-1})$~Mpc, the random error component
being largely due to intrinsic scatter in the FC locus. This
result is shown in Figure~\ref{f02}. Note that recent evidence
that the Fundamental `plane' becomes a curved sheet for normal
[i.e. low surface brightness] dwarf ellipticals
\citep{GG 2003,G 2005} may reduce the accuracy of the above
calculation but does not invalidate it. This is because the
curvature is such that should M32's true distance be similar to
that of M31, it would be even more deviant from the low-luminosity
end of the Fundamental `sheet' than it already is from the
low-luminosity linear extrapolation of the plane as shown in
Figure~\ref{f02}.  Also, should M32's true distance be of the
order of 2.3~Mpc, it would no longer be a dwarf galaxy and could
therefore be expected to conform to the essentially planar part of
the Fundamental plane for early-type giant and intermediate
galaxies.

FC-galaxy \citet{S 1968} parameters were derived from the surface
photometry of \citet{CB 1987} and \citet{CCDo 1994}\footnote{There
were 44 galaxies in this combined sample, but NGC~1428 is now
known to be affected by a bright foreground star \citep{M+ 1998}
and therefore had to be excluded from the analysis.} following
procedures (and nomenclature) previously used by \citet{Y 2001}
for Virgo photometry.  As \citet{S 1968} parameters are seeing
dependent (\citealt{Y 2004} and references therein) and
\citet{CB 1987} did not quote seeing discs for their FC
photometry, we estimated their seeing by fitting Gaussians to the
nuclei of the 9 nucleated dwarf ellipticals in their sample. The
overall mean seeing full-width half maximum (FWHM) for both
sources of photometry was subsequently found to be 1.8~arcsec. A
more detailed description of these procedures will be given by
Young et al.\ (in preparation).

Computation of M32's global parameters involved reconstructing a
1-arcsec pixel two-dimensional image of the galaxy from the
surface photometry of \citet{K 1987}. We transformed the
$r_{TG}$-band photometry to $B$-band photometry according to
$B=r_{TG}+1.21$~mag and evaluated $\mu_{B}$ for each pixel by
fitting a cubic spline to points on the straight line that passed
through both the image centroid and the centre of the pixel
concerned. The points fitted were the intersections between the
radial line and the published elliptical isophotes taking full
account of the differing position angles of each isophote. For all
pixels exterior to the outermost isophote, $\mu_B$ values were
estimated by linear extrapolation in $\mu_{B}$-$r$ space. The
synthetic image was then scale reduced to a series of images
corresponding to the angular sizes that would be subtended if the
galaxy's actual distance were 0.76 through 4.0~Mpc, and it were
subsequently relocated to 20~Mpc. This scale reduction was
accompanied by simultaneous re-gridding in order to create
1/3-arcsec pixel images, which were then convolved with
point-spread functions (modelled using Equation~6 of
\citealt{R 1996} which has a dynamic range of 14~mag~arcsec$^{-2}$)
in order to mimic the effects of a seeing disc of 1.8-arcsec FWHM.
Surface photometry was then performed and \citet{S 1968} models
fitted to the final profiles excluding isophotes corresponding to
$r<0.9$~arcsec in each final image and to $r>220$~arcsec in the
original image. Extrapolated regions exterior to the outermost
isophote of the published surface photometry therefore underwent
convolution but were not fitted by the profile models.

\begin{figure}[!ht]
\centering
\includegraphics[height=85mm,angle=-90]{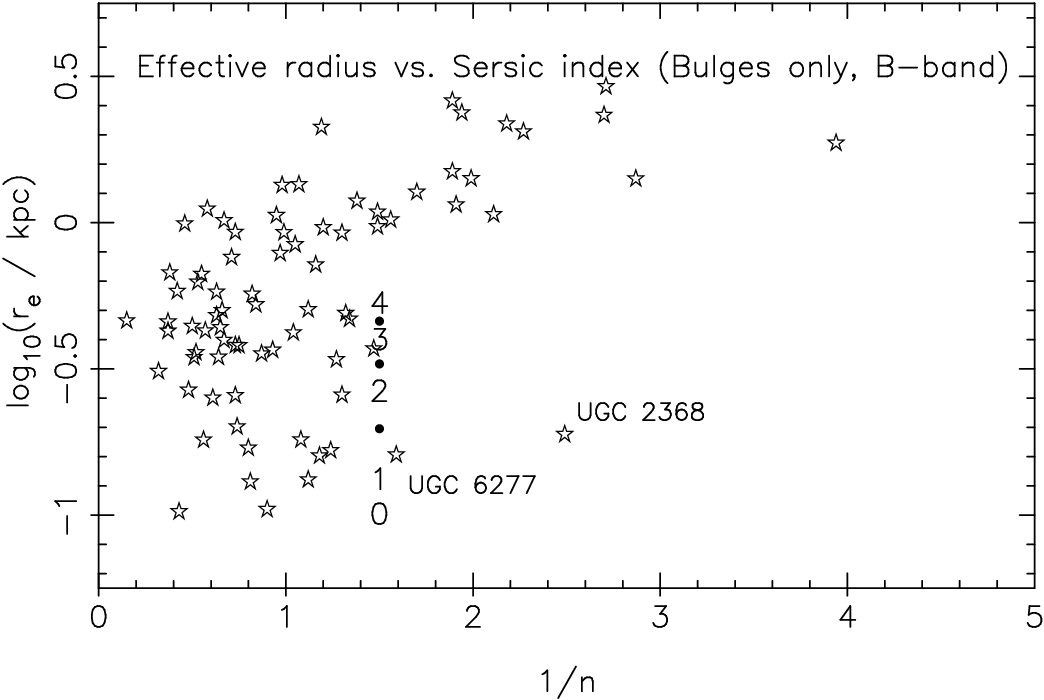}
\vspace{-4mm} \caption{\baselineskip 3.6mm Expected positions of
M32's bulge as a function of its true distance (which is unknown)
relative to the bulges of those 86 spiral galaxies in the sample
of \cite{DjVdk 1994} with $B$-band data. Parameters for M32 are
from \cite{G 2002} without further seeing corrections applied,
whilst all \cite{S 1968} indices ($n$), effective radii
($r_{e}$/arcsec) and radial velocities for the spiral bulges are
from the addendum to \cite{G 2001} as opposed to the original
paper. Note that $n$ in our notation is equivalent to $1/n$ in the
notation of \cite{G 2001} and \cite{G 2002}. Some of the
[significant] scatter in the locus defined by the spiral bulges
must be due to the sample galaxies being at widely differing
distances that could only be estimated approximately--via their
radial velocities (assuming $H_{0} = 75$~km~s$^{-1}$~Mpc$^{-1}$)
and with corrections for Virgocentric infall from \cite{M+ 2000}.
However, uncertainties in \cite{S 1968} profile parameters are
also quite significant for some galaxies. The notation for M32
follows Fig.~\ref{f02} whilst spiral bulges are denoted by $\star$
symbols. }
\label{f03}
\end{figure}

So far we have compared the global properties of M32 and
early-type galaxies in the FC. Should M32 be a lenticular galaxy,
i.e.\ a two-component system, as first proposed by \citet{G 2002},
we should be able to refine our comparisons by comparing
bulge-specific properties after profile de-composition. Now,
\citet{G 2002} finds the \citet{S 1968} index of M32's bulge to be
$n=0.67$ in our notation, and has already demonstrated that many
of M32's properties are characteristic of normal lenticular
systems e.g.\ (1) its luminosity excess at intermediate radial
distances with respect to a single component \citet{S 1968}
luminosity-profile model, (2) its significant ellipticity gradient
(suggesting the increasing dominance of a disc at high radial
distances) as well as (3) its bulge-to-disc size and luminosity
ratios. In the present study though, our main interest is in
distance-dependent parameters, notably the effective radii
($r_{e}$) of the bulges of both M32 and known two-component
systems. As evident from Figure~\ref{f03}, the small effective
radius of M32's bulge (which \citealt{G 2002} found to be
27~arcsec or $\sim$100~pc if M32's distance is similar to that of
M31) is highly consistent with it being a background galaxy.

\end{document}